%% file: main.tex
\documentclass{article}
\PassOptionsToPackage{numbers,compress}{natbib}
\usepackage[final]{neurips_2025}

\usepackage{times}
\usepackage{booktabs}
\usepackage{caption}
\usepackage{enumitem}
\usepackage[group-separator={,},group-minimum-digits=3]{siunitx} 
\usepackage{graphicx}
\usepackage{tikz}
\usepackage{multirow}
\usepackage{tabularx}
\usepackage{listings}
\usepackage{wrapfig}
\usepackage{float}
\usepackage{pifont}

\usepackage[utf8]{inputenc} 
\usepackage[T1]{fontenc}    
\usepackage{url}            
\usepackage{amsfonts}       
\usepackage{nicefrac}       
\usepackage{microtype}      
\usepackage{xcolor}         

\usepackage{tcolorbox}

\usepackage{xurl}

\tcbuselibrary{skins,breakable}

\definecolor{ForestGreen}{RGB}{27, 158, 119}  
\definecolor{BurntOrange}{RGB}{217, 95, 2}    

\usepackage{tcolorbox}
\usepackage{adjustbox}
\tcbuselibrary{breakable, skins}

\usepackage{url}
\usepackage[breaklinks=true,backref=page,colorlinks=true,linkcolor=black,citecolor=black,urlcolor=blue]{hyperref}

\let\oldhref\href
\renewcommand{\href}[2]{\oldhref{#1}{\underline{#2}}}

\title{\benchmark: Evaluating Language Models on Cloud Infrastructure Code}

\author{
\textbf{Natalia Tarasova} \quad Enrique Balp-Straffon \quad \textbf{Aleksei Iancheruk} \quad \textbf{Yevhenii Sielskyi} \quad \\ \textbf{Nikita Kozodoi} \quad \textbf{Liam H. Byrne} \quad \textbf{Jack Butler} \quad \textbf{Dayuan Jiang} \quad  \textbf{Marcin Czelej} \quad \\ \textbf{Andrew Ang} \quad \textbf{Yash Shah} \quad \textbf{Roi Blanco} \quad \textbf{Sergei Ivanov}
\\[0.2cm]
Amazon Web Services \quad 
}

\newcommand{\benchmark}{SWE-InfraBench}

\begin{document}

\maketitle

\input{section/abstract}
\input{section/introduction}

\input{section/related}

\input{section/benchmark}

\input{section/results}

\input{section/discussion}

\newpage

\bibliographystyle{splncs04}
\bibliography{bibliography}

\newpage
\appendix
\section*{Appendix}
\input{appx/statistics}
\input{appx/cdk_versions}
\input{appx/source-licences}

\input{appx/tasks_teaser}
\input{appx/prompts}
\input{appx/model_configs}
\input{appx/verbosity}
\input{appx/performance_by_donor_type}
\input{appx/consistency_analysis}
\input{appx/error-distribution}

\end{document}

%% file: section/abstract.tex
\begin{abstract}
Building infrastructure-as-code (IaC) in cloud computing is a critical task, underpinning the reliability, scalability, and security of modern software systems. Despite the remarkable progress of large language models (LLMs) in software engineering -- demonstrated across many dedicated benchmarks -- their capabilities in developing IaC remain underexplored. Unlike existing IaC benchmarks that predominantly center on declarative paradigms such as Terraform and involve generating entire codebases from scratch, our benchmark reflects the incremental code edits common in enterprise development with imperative tools like the AWS CDK. We present \benchmark{}, a diverse evaluation dataset sourced from dozens of real-world IaC codebases that challenge LLMs to perform realistic code modifications in AWS CDK repositories. Each example requires models to implement changes to existing codebases based on natural language instructions, with success determined by passing provided test cases. These tasks demand sophisticated reasoning about cloud resource dependencies and implementation patterns beyond conventional code generation challenges.
Our evaluation results reveal significant limitations in current LLMs showing that even state-of-the-art systems struggle with many tasks -- the best model, Sonnet 3.7, succeeds in only 34\% of cases, while specialized reasoning models like DeepSeek R1 achieve just 24\% success. The \benchmark{} dataset is available on \href{https://www.kaggle.com/datasets/64e59070fd51c0278560b01eb5dc4f3c447d5268cdabe5a350d2969e4413fea5}{Kaggle}.


\end{abstract}

%% file: section/introduction.tex
\section{Introduction}
\label{section:introduction}

Modern cloud computing platforms enable unprecedented scalability and automation, but realizing its potential requires managing highly complex infrastructure configurations \citep{Canalys2025, EdgeDelta2025}. Infrastructure as Code (IaC) has emerged as a DevOps practice to meet this need, treating the specification of cloud resources as software artifacts \citep{Quattrocchi2023, Pahl2025, Morris2020}. With IaC, cloud environments (servers, networks, etc.) are defined and provisioned using code, bringing software engineering rigor (version control, code review, continuous integration) to infrastructure management. In broad terms, IaC approaches fall into two paradigms: declarative and imperative. A declarative IaC tool describes the desired end state of the infrastructure, leaving it to the system to “figure out” the necessary actions. An imperative IaC tool, in contrast, specifies the exact sequence of commands to reach that end state \citep{KuduLab2023, Larssen2024, Wicher2025}. 

Two dominant frameworks exemplify these paradigms in today’s cloud IaC landscape: Terraform and AWS Cloud Development Kit (CDK). Terraform is a widely-adopted declarative IaC tool that uses a domain-specific language (HCL) to let operators specify the final state of infrastructure resources \citep{Buchh2023}. While this approach has strengths (platform-agnostic abstractions, a robust planning mechanism), it also has notable limitations for complex, evolving systems. Purely declarative configurations can become brittle -- small changes may require non-intuitive refactoring or state file surgery -- and inflexible, as HCL lacks the expressive power of general-purpose languages for handling conditional logic, loops, or abstractions. 

The AWS CDK directly addresses these pain points by taking an imperative developer-friendly approach \citep{aws_cdk_developer_guide, godaddy_cdk_case_study, sbes}. CDK allows developers to define cloud infrastructure using familiar programming languages (TypeScript, Python, etc.), leveraging the full power of those languages to encode loops, reuse components, and incorporate conditionals. This means infrastructure definitions can be developed in the same codebase and environment as the application code, with complete IDE support and testing frameworks \citep{Poccia2019, Wiggers2019, Poccia2019}. The two philosophies yield different developer experiences: Terraform favors stability but limits agility, while CDK supports incremental change and continuous delivery. Since industrial software development often demands iterative experimentation and close coupling with application logic, we adopt AWS CDK for this benchmark, which enables incremental and testable infrastructure. 

Surprisingly despite widespread industrial adoption of IaC frameworks, most of the coding evaluation benchmarks for LLMs focused on programming tasks such as code completion and translation \citep{jimenez2023swe, chowdhury2024swebench, rashid2025swe, du2024classevalt, pan2024lost}, debugging and unit test synthesis \citep{prasad2025learning, rahman2025utfix, jiang2024survey, etsenake2024understanding, raihan2024code}, but not on IaC generation. Evaluation of LLMs specifically for imperative infrastructure-as-code editing tasks remains a substantially underexplored research area due to unique challenges associated with interpreting and maintaining consistency across large stateful codebases, as well as the inherent difficulty of curating relevant examples that require specialized in-domain knowledge. 


\input{figures/teaser}

To address the gap in IaC evaluation, we introduce \benchmark{}, a benchmark dataset designed to assess language models' capabilities on IaC tasks within the AWS CDK framework. \benchmark{} comprises 100 diverse examples, each containing complete CDK repositories, modification instructions reflecting real-world development scenarios, and multiple unit tests verifying the correctness of LLM-generated implementations. To the best of our knowledge, \benchmark{} is the first specialized benchmark for evaluating language models on CDK editing tasks, distinguishing itself from existing IaC benchmarks \citep{kon2024iac, xu2024cloudeval} in two critical dimensions. First, previous benchmarks primarily target Terraform and its declarative syntax, whereas our benchmark leverages the imperative programming languages used by CDK, a framework widely adopted in industry for infrastructure development. Second, \benchmark{} emphasizes incremental codebase modifications rather than complete application synthesis, reflecting the iterative development patterns predominant among cloud practitioners in enterprise environments.

Our work makes the following contributions:
\begin{enumerate}
    \item We provide a new collection of a hundred instruction-based tasks for modification of IaC codebases that are challenging to modern LLMs, collected from 34 real-world assets. Each task includes a full AWS CDK project with a specific modification instruction and a suite of unit tests that must pass after the change. The dataset is available on \href{https://www.kaggle.com/datasets/64e59070fd51c0278560b01eb5dc4f3c447d5268cdabe5a350d2969e4413fea5}{Kaggle}.
    \item We introduce a systematic LLM-driven pipeline for generating and validating new benchmark examples from arbitrary CDK repositories. This approach enables anyone to create realistic tasks from any IaC project and ensures each example is valid by verifying that the modified code passes all tests after synthesis.
    \item We conduct a comprehensive experimental analysis across diverse LLM architectures, including both proprietary and open-source models with varying parameter scales and architectural designs, establishing baseline performance metrics for this domain. Going beyond individual LLM models, we evaluate performance of multi-turned agents enhanced with error messages, test results, and RAG on IaC documentation.
\end{enumerate}

%% file: figures/teaser.tex
\begin{figure}
    \centering
    \includegraphics[width=0.85\textwidth]{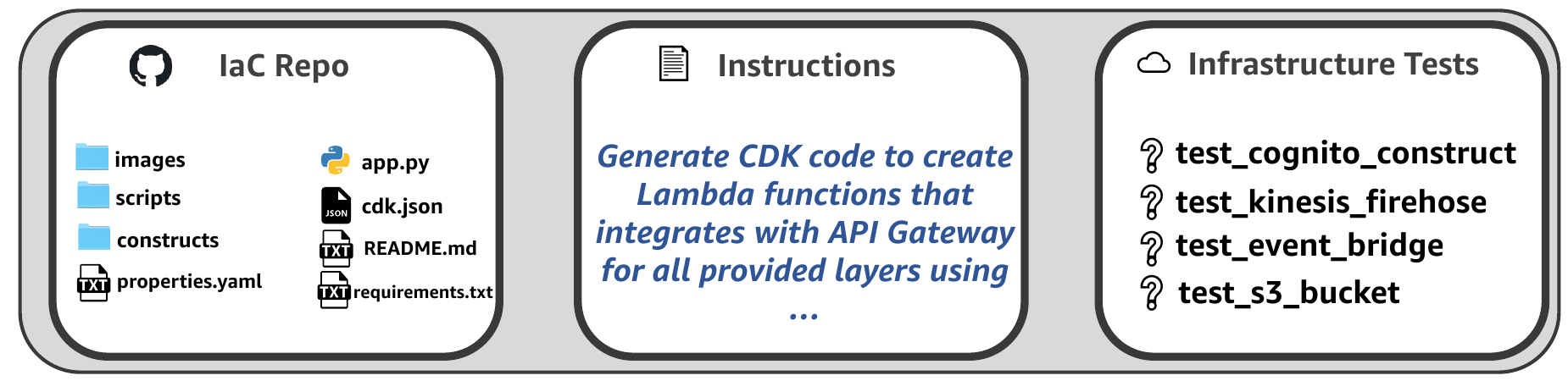}
    \caption{
    Each \benchmark{} task includes an IaC repository, natural language modification instructions, and automated tests verifying the correctness and integration of the generated solution.
    }
    \label{fig:teaser}
\end{figure}

%% file: section/related.tex
\section{Related Work}
\label{sec:related-work}
\textbf{LLMs for Infrastructure-as-Code Generation}

IaC has become crucial for modern DevOps and MLOps, enabling programmatic management of cloud resources. Research has explored using LLMs to automate IaC generation. Early work investigated generating Ansible YAML from natural language \citep{kawaguchi2022implementation, pujar2023automated}. However, recent evaluations reveal LLMs perform significantly worse on IaC tasks compared to general Python generation \citep{kon2024iac}, highlighting limitations in domain-specific configuration automation. Benchmarks like CloudEval-YAML \citep{xu2024cloudeval} and IaC-Eval \citep{kon2024iac} address this gap, with IaC-Eval showing GPT-4 achieving only 19.36\% pass@1 accuracy on human-curated IaC scenarios, despite scoring 86.6\% on EvalPlus. These benchmarks focus on declarative frameworks such as Terraform and ask to generate entire codebase from prompt instructions, which is different from our \benchmark{} that targets imperative infrastructure code modifications within AWS CDK repositories. Finally, the dataset LIG-MM \citep{liu2024ligmm} explores symbolic and neural methods to address the formal reasoning challenges in IaC tasks.

\textbf{Code Generation and Editing Benchmarks} Advancements in LLM code generation evaluation have created more comprehensive benchmarks spanning from single functions to multi-file problems. Meanwhile, research has expanded code editing benchmarks beyond traditional bug fixing to address additional tasks like code completion. HumanEval~\citep{chen2021evaluating} remains a popular benchmark for evaluating code generation, with recent extensions targeting multilingual~\citep{athiwaratkun2022multi}, editing~\citep{yu2024codereval}, and reasoning variants~\citep{muennighoff2023octopack}. However, these tasks often focus on short, self-contained functions. CrossCodeEval~\citep{ding2023crosscodeeval} addresses this limitation by introducing multi-file completion problems that require reasoning across files and navigating real codebases -- bringing code benchmarks closer to industrial development settings. In order to better simulate real world engineering tasks on identifying location where the code editing is required, SWE-PolyBench, \citep{rashid2025swe} was introduced as an incremental development to the SWE-bench \citep{jimenez2023swe} as the first dataset that mirrors real-world software engineering specifically covering over 2000 curated coding challenges across Java, JavaScript, TypeScript and Python along with evaluation metrics like live level localization and concrete Syntax Tree node level retrieval. Similarly, Mercury~\citep{du2024mercury} evaluates models not only on correctness but also on runtime performance, highlighting inefficiencies in LLM-generated code limiting real-world utility. In code editing, the focus has largely been on bug fixing, with fewer benchmarks addressing tasks like code completion. Initiatives like CodeEditorBench \citep{guo2024codeeditorbench}and CanItEdit \citep{cassano2023can} have aimed to fill this gap. CodeEditorBench tried to address this gap, by creating a dataset of tasks to help assess the performance of LLMs in tasks like code debugging, code translation, code requirement switching and even code polishing. But the use of competitive programing problems meant that the real-world replicability of the performance was not tenable. CanItEdit, introduces a dataset of instructional editing problems with a novel ExcessCode metric~\citep{guo2024codeeditorbench,cassano2023can}.

%% file: section/benchmark.tex
  \section{\benchmark{}}
\label{sec:benchmark}

\input{figures/dataset-pipeline}

\benchmark{} consists of a collection of IaC codebases implemented using AWS CDK. Each codebase is paired with a prompt that contains a natural language instruction to generate new code blocks extending the existing infrastructure. For each codebase, we also provide multiple unit tests that enable verification of potential solutions. The task for an LLM is to produce a code change that satisfies the requirements outlined in the prompt and passes the associated tests. On average, one codebase includes 10 files spanning 30 thousand characters that are provided to the LLM as context. Further statistics of the dataset tasks can be found in Appendix ~\ref{sec:benchmark:features-of-benchmark} and ~\ref{app:cdk_versions}.

CDK code is synthesized into AWS CloudFormation~\citep{aws_cloudformation_guide} templates -- JSON or YAML files that describe the desired AWS resources and their relationships. This process enables automated testing of infrastructure definitions without actual deployment, which is particularly valuable since deploying cloud infrastructure can be time-consuming and resource-intensive. This makes CDK a good candidate for benchmarking LLMs on cloud infrastructure generation tasks.

\subsection{Benchmark Construction}
\label{sec:benchmark:construction}

Creating a high-quality IaC dataset requires identifying precise, robust examples with clear objectives and verifiable outcomes. To achieve this at scale and with consistent quality, we developed a multi-stage pipeline combining human expertise with LLMs assistance, which is shown in Figure~\ref{fig:dataset-pipeline}.

\textbf{Stage I: Donor Repository Selection and Task Generation.}
We begin by identifying high-quality source repositories implementing IaC. Our benchmark draws from both open-source repositories and custom-developed sources, with licensing details provided in Appendix~\ref{app:source-licences}. Using these repositories, an LLM (typically Claude Sonnet 3.5 or 3.7) analyzes the codebase and generates initial task suggestions. Each task includes a code block that should be masked (also referred to as the canonical solution), a prompt with functional requirements, and unit tests to verify correctness of generated solutions. Human domain experts then review these suggested tasks and select the most promising candidates.

\textbf{Stage II: Critique and Refinement.}
Each task candidate undergoes either manual refinement or an automated improvement process. For automated refinement, we use a critic-based approach:

1. \textit{Validate}: The system validates the tests by running them on the canonical solution (masked code) to ensure they are passed by a correct implementation of the required functionality.

2. \textit{Critique}: Two different critic models (Sonnet 3.7, o3, or o4-mini), which are distinct from the generator, perform a comprehensive evaluation of the task suggestion across four key dimensions:
   
\begin{itemize}
    \item  \textit{Prompt vs. Functional Requirements Alignment}: Evaluates whether the prompt accurately describes all necessary functional aspects of the masked code, is sufficiently detailed for correct implementation, avoids ambiguities, and balances completeness with avoiding implementation details.
    \item  \textit{Implementation Generality}: Assesses whether the tests allow for multiple valid implementation approaches that satisfy the functional requirements, rather than enforcing a specific implementation pattern.
    \item  \textit{Test Quality}: Analyzes each individual test to determine its purpose, how it relates to prompt requirements, and whether it is appropriately written to verify the requirement without being overly restrictive.
    \item  \textit{Test Completeness}: Verifies that the tests collectively cover all requirements mentioned in the prompt, including edge cases and error conditions.
\end{itemize}

        The second critic model is only evaluated if the first one was satisfied. Critics provide detailed feedback with specific corrections and code improvements. Both critics must approve all four dimensions for the suggestion to pass. The full critic prompt template used for evaluation is provided in Appendix~\ref{app:critic_prompt}.

3. \textit{Improve}: Based on validation results and critics' feedback, the generator LLM refines the suggested task, addressing any identified issues in the prompt or tests. The prompt for the generator model is provided in Appendix~\ref{app:generator_prompt}.

Throughout this process, human experts can guide the refinement by providing specific instructions to both the generator and critics, allowing for targeted improvements while maintaining the benefits of automated evaluation.

\textbf{Stage III: Expert Review and Finalization.}
Human experts review and perform optional final modifications in each task to ensure:
\begin{itemize}
    \item \textbf{Fairness}: The prompt and tests are properly aligned so any valid solution following the prompt will pass the tests;
    \item \textbf{Test coverage}: Tests adequately verify the prompt requirements;
    \item \textbf{Challenge level}: The task presents an appropriate difficulty for state-of-the-art models;
    \item \textbf{Automatic validation}: The canonical solution passes all tests.
\end{itemize}

This multi-faceted approach ensures that our benchmark contains systematically vetted and validated infrastructure-as-code tasks that provide a rigorous evaluation framework for LLMs.

The final benchmark tasks are stored as JSON files containing the context files (with masking tags), natural-language prompts, canonical solutions, tests, and CDK version information. See Appendix~\ref{subsection:tasks-example-teaser} for a task format preview and~\ref{subsection:tasks-prompt-examples} for example prompts.

\subsection{Dataset Collection Challenges}
\label{sec:benchmark:challenges}

Building a high-quality IaC benchmark required substantial engineering expertise. Each task needed rigorous human oversight to ensure it was challenging, objective and fair, with aligned prompts, tests, and solutions. This process demanded experienced IaC developers with deep cloud knowledge. On average, an engineer could produce five high-quality dataset items per day, which required us to employ tens of cloud practitioners collaborating on building the dataset.

As outlined in Section \ref{sec:benchmark:construction}, we developed a semi-automated pipeline using LLMs to jumpstart creation, provide refinement, and offer critique. While this helped to derive examples from existing repositories, human oversight remained essential for the final validation and quality control. A major obstacle was the lack of test coverage in open-source IaC repositories. Most open-source IaC projects lack comprehensive tests for CloudFormation templates, requiring us to develop tests for each task ourselves.

By releasing our task generation pipeline alongside the benchmark, we aim to establish a scalable process that directs the focus of the human expertise on the area where it matters most: refining and validating machine-generated tasks rather than creating new items from scratch.

\subsection{Problem Definition} 
\label{sec:benchmark:problem-definition}


\textbf{Model Input.}
Given a textual instruction describing a desired functionality within the IaC codebase, a language model is tasked with generating appropriate code modifications that fulfill the described objective. Specifically, the model receives a detailed prompt outlining the components, their interactions, and constraints. For example, a prompt might instruct the model to set up an event-driven pipeline involving components such as a message broker, data delivery stream, object storage bucket, access control policies, and event routing rules. Examples of the prompts can be found in Appendix~\ref{subsection:tasks-prompt-examples}. The entire codebase is provided to the LLM as additional context. The resulting generated solution is represented as structured code edits that can be directly integrated into the existing codebase.

\textbf{Evaluation Metrics.}
To objectively evaluate the correctness of the generated solutions, we integrate generated code modifications into the target codebase and subsequently execute predefined unit tests designed to validate the implemented functionality. These tests check for the presence and correctness of the infrastructure components described by the prompt, verifying criteria such as resource creation, correct permissions and configurations, and successful integration between components. A solution is considered correct if it integrates into the codebase without errors and passes all associated unit tests. 

The key evaluation metrics for our benchmark vary according to the number of trials for each model on each task. The considered metrics are:

\begin{itemize}
    \item \textbf{pass@k} as defined in \citep{chen2021evaluating}: Probability that LLM completes a task at least once in $k$ trials.
    \item \textbf{Correctness}: Average number of tasks completed (equals to pass@1 for one-trial experiments). A task is completed if all of its tests are passed.
    \item \textbf{Generation Success}: Proportion of generated solutions that follow the required format and could be integrated within existing codebase.
    \item \textbf{Passed Tests Share}: Proportion of tests across all examples that are passed.
\end{itemize}

%% file: figures/dataset-pipeline.tex
\begin{figure}[h]
\centering
\includegraphics[width=1\linewidth]{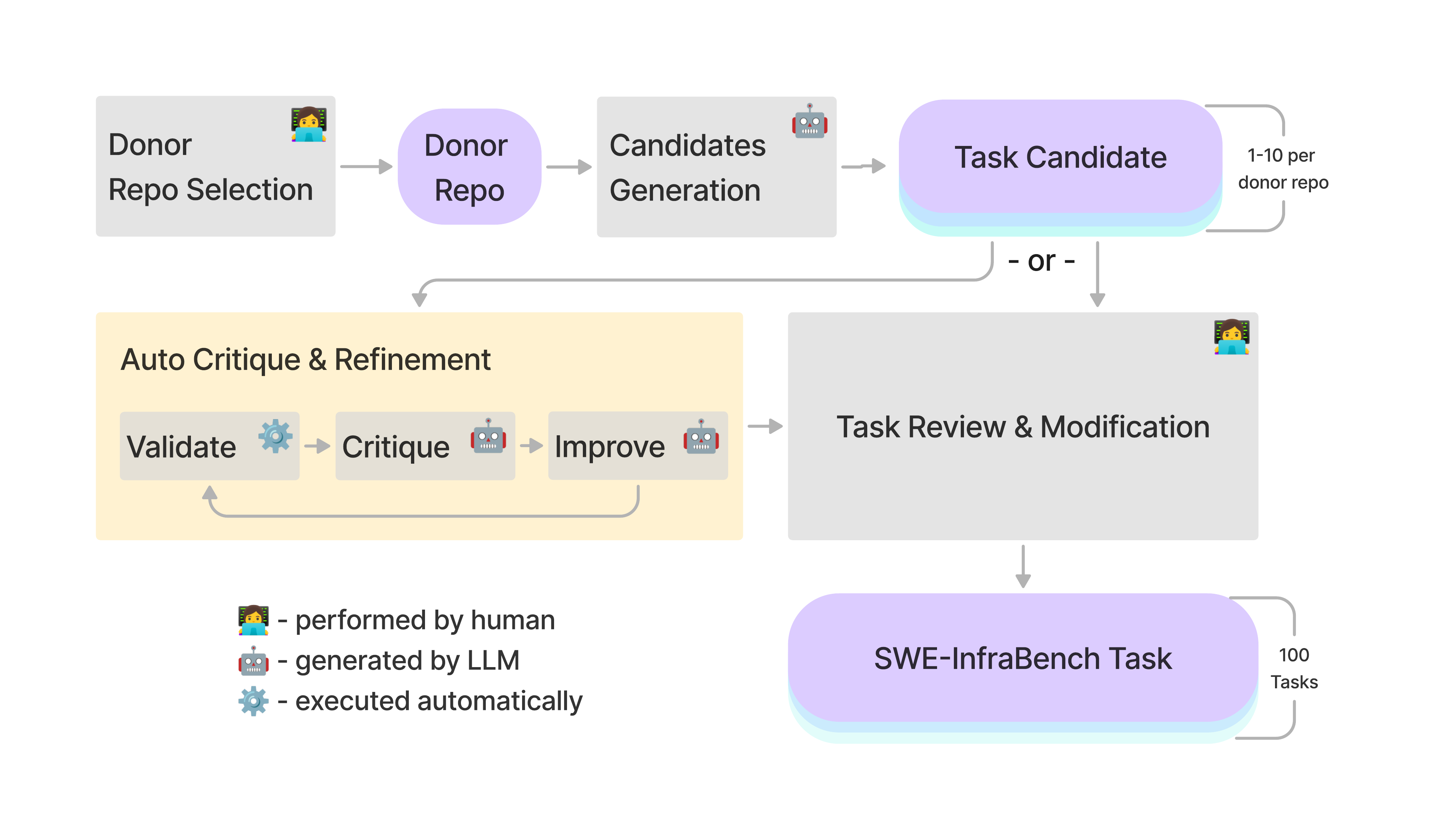}
\caption{
\benchmark{} task instances are created from open-source and custom developed IaC repositories in a semi-automated manner with a rigorous human engineers oversight.
}
\label{fig:dataset-pipeline}
\end{figure}

%% file: section/results.tex
\section{Experimental Results}
\label{sec:results}

We evaluate 20 state-of-the-art language models on 100 \benchmark{} tasks. Each model is given a single attempt to generate a solution for each task. The solutions are integrated into the task context repository, and tests are run on the resulting CloudFormation output to assess correctness.

\benchmark{} proves to be challenging, with even the top-performing models solving less than 35\% of the tasks completely. This highlights the complexity of IaC generation, which requires both domain knowledge, precise syntax adherence as well as deep prompt and context understanding.

\input{tables/zero-shot-general-results}

Table~\ref{tab:model-metrics} shows the results for 20 models on the metrics defined in Section~\ref{sec:benchmark:problem-definition}, averaged over all code generation tasks. The best performance is achieved by the Claude Sonnet family of models (3.7, 3.5 V2 and 3.5). While these models reportedly excel in code generation benchmarks \citep{anthropic2025claude37}, such as SWE-bench \citep{jimenez2023swe}, it is also worth noting that the input prompt format used for all the Anthropic models incorporates additional XML tags, unlike the default prompt template for other LLMs (see Appendix~\ref{subsection:direct-solver-prompts}). Notably, reasoning models such as Gemini 2.5 Pro, DeepSeek R1, OpenAI o3, and o4-mini demonstrate stronger capabilities than non-reasoning models such as GPT-4.1, Claude 3 Haiku, and Gemini 2.0 Flash. This aligns with our expectations, as IaC generation requires substantial reasoning to understand the requirements, determine appropriate resources, and establish correct relationships between components. Nevertheless, as shown in Appendix~\ref{section:model-configs}, reasoning models require an increased token budget.

Models generally achieve high generation success rates (over 94\%), indicating they could follow the output format, but their ability to produce fully correct solutions varies significantly. This suggests the challenge is not understanding the requirements but implementing the correct functionality. The "Passed Tests Share" metric shows that even when models do not solve tasks completely, they often implement substantial portions of the code correctly.

Since a substantial share of the tasks are created from open-source data, we separately confirm that tasks originating from open-source repositories are not less challenging than custom-developed ones. The details of this analysis are provided in Appendix~\ref{section:performance-by-donor}.

\subsection{Consistency Analysis}

To evaluate models reliability, we conduct a separate multi-trial analysis on a selected subset of models. For each task, we execute five independent solution attempts (trials) per model, allowing us to measure both peak performance and consistency. This approach addresses two critical questions: (1) can a model solve a given task at least once in K attempts (pass@K), and (2) how consistently does the model produce the same outcome across attempts?

As shown in Table~\ref{tab:pass-at-k-metrics}, higher pass@k values indicate that models can solve more tasks given multiple attempts, which is valuable for iterative development scenarios. At the same time, average correctness across five generations for the best-performing models falls below 30\%, indicating that it is challenging to achieve consistently accurate generations. These results indicate that in practical applications, multiple generation attempts can increase success rates, particularly for complex IaC tasks where small syntax variations can determine the successful infrastructure creation. We provide further analysis of the relationship between the average correctness and success consistency in Appendix~\ref{section:consistency-analysis}. 

\input{tables/success_k}

\subsection{Error Type Distribution}

Our analysis of error types across 20 LLMs reveals distinct patterns in their ability to generate IaC solutions, as shown in Figure~\ref{fig:error-dist}. Incorrect property usage in CDK constructs and other syntax errors make a considerable part of all failures. Such issues can be caused both by models lacking information about particular components syntax as well as models having a knowledge cutoff that is earlier than more recent CDK versions.

\input{figures/error-distribution}

These findings suggest that agent-like approaches that allow models to make an attempt to generate the code, analyze the errors, access recent documentation and refine solutions could substantially improve performance, particularly for syntax-related issues. This would better mirror the iterative development process that human developers use when working with IaC problems.

\subsection{Multi-Turn Agent Performance}

To investigate the potential benefits of iterative approaches, we implement a two-turn agent that provides models with feedback from their initial attempts. When a model fails to solve a task on the first try, error messages and test results are fed back to the model for a second attempt. The feedback mechanism is implemented with two distinct verbosity configurations: low-verbosity ($V_L$), which provides basic error messages and test pass/fail counts, and high-verbosity ($V_H$), which includes comprehensive output from tests with full tracebacks and detailed diagnostic information (see Table~\ref{tab:verbosity-configurations} for a complete specification).

Additionally, we enhance this approach with a Retrieval-Augmented Generation (RAG) system that operates in a two-phase process. In the first phase, along with generating the initial solution, models produce relevant keywords for documentation retrieval. These keywords are used to fetch the top-k most relevant documentation pages from AWS documentation. If the initial solution fails, the second phase incorporates both the error feedback and the retrieved documentation into the context for the model's second attempt.

Our results demonstrate that multi-turn interaction substantially improves performance across all models, as shown in Table~\ref{tab:two-turn-comparison}. Claude 3.7 Sonnet achieves the highest correctness score of 64\% with the high-verbosity configuration (among standard two-turn configurations), representing a significant absolute improvement of 30 percentage points over its one-turn baseline. Additional details on error distribution changes for multi-turn approach can be found in Appendix~\ref{section:error-dist}.

The interaction between model scale and verbosity reveals other interesting patterns. While larger models demonstrate substantial improvements with increased verbosity, smaller models like Mistral Large show more modest gains (two percentage points gain in correctness score for standard configuration when going from $V_L$ to $V_H$). This pattern indicates that the ability to effectively utilize detailed error information may be contingent on model capacity.

\input{tables/two-turn-general-improvement-and-rag}

The verbosity effect proves particularly noteworthy, with $V_H$ configurations consistently outperforming $V_L$ across LLMs, suggesting that detailed diagnostic information enables more effective error correction. RAG configurations add further complexity to the performance landscape, with Claude 3.5 Sonnet V2 achieving the highest correctness score of 65\% -- a substantial 33 percentage points improvement over its one-turn performance. The effectiveness of RAG varies significantly across model architectures. Smaller models like LLaMA 4 Maverick show modest but consistent improvements with RAG, while some high-performing models like GPT-4.1 demonstrate reduced effectiveness (26\%) compared to their standard two-turn performance (48\%). This heterogeneity suggests that RAG implementation may require model-specific optimization strategies.


These results warrant several important considerations. First, full test failure tracebacks may reveal details that might inadvertently help models correct their solutions without genuine understanding of the underlying problems. The error messages often contain specific values expected for CloudFormation properties, potentially allowing models to pattern-match rather than reason about the proper solution. This may make the observed performance gains overoptimistic with regards to real-world use cases, where the true CloudFormation templates are unknown. Second, we observe that DeepSeek R1 allocates a substantial number of tokens to reasoning during its second attempt, which requires the raise of maximum output token parameter to prevent it from exhausting the token limit before generating a complete solution. Finally, the multi-turn approach, while effective, comes with increased computational costs, both in terms of token consumption and inference time.

Our results suggest that enabling iterative refinement with detailed feedback is a promising direction for enhancing model capabilities. The consistent advantage of high-verbosity configurations highlights the importance of detailed diagnostic information in enabling successful iterative refinement, particularly for more sophisticated model architectures. However, the effectiveness of enhancement strategies like RAG is not uniform across model architectures.

%% file: tables/zero-shot-general-results.tex
\begin{table}
\centering
\caption{Performance metrics for proprietary and open-source LLMs on IaC generation tasks. Models are grouped by source type and provider, and sorted from the most recent to the oldest variant within each group. \textsuperscript{\dag} indicates LLMs executed with reasoning capabilities. \textsuperscript{\ding{61}} Claude 3.7 was executed without extended reasoning. Top results are in \textbf{bold}.}
\label{tab:model-metrics}
\resizebox{\textwidth}{!}{
\begin{tabular}{lllccc}
\toprule
\textbf{Source} & \textbf{Company} & \textbf{Model} & \textbf{Correctness} & \textbf{Generation Success} & \textbf{Passed Tests Share} \\
\midrule
\multirow{11}{*}{\rotatebox[origin=c]{90}{\textbf{Proprietary}}} 
& Anthropic & Claude 3.7 Sonnet \textsuperscript{\ding{61}} & \textbf{34\%} & \textbf{100\%} & \textbf{53.1\%} \\
&  & Claude 3.5 Sonnet V2 & 32\% & \textbf{100\%} & 47\% \\
&  & Claude 3.5 Sonnet & 29\% & \textbf{100\%} & 47.9\% \\
&  & Claude 3 Haiku & 8\% & 88\% & 10.7\% \\
\cmidrule(lr){2-6}
& Google & Gemini 2.5 Pro (03-25 Preview) \textsuperscript{\dag} & 29\% & 97\% & 42.5\% \\
&  & Gemini 2.5 Pro (05-06 I/O Edition Preview) \textsuperscript{\dag} & 29\% & 95\% & 41.1\% \\
&  & Gemini 2.0 Flash Lite & 5\% & 79\% & 11.5\% \\
&  & Gemini 2.0 Flash & 4\% & 43\% & 6.9\% \\
\cmidrule(lr){2-6}
& OpenAI & OpenAI o3 \textsuperscript{\dag} & 23\% & \textbf{100\%} & 30.8\% \\
&  & OpenAI o4-mini \textsuperscript{\dag} & 23\% & 99\% & 32.1\% \\
&  & GPT-4.1 & 18\% & \textbf{100\%} & 26.4\% \\
&  & GPT-4o Mini & 4\% & 84\% & 7\% \\
\midrule
\multirow{9}{*}{\rotatebox[origin=c]{90}{\textbf{Open-Source}}}
& DeepSeek & DeepSeek R1 \textsuperscript{\dag} & 24\% & \textbf{100\%} & 34.9\% \\
\cmidrule(lr){2-6}
& Mistral & Mistral Large & 14\% & 89\% & 20.3\% \\
&  & Codestral & 9\% & 96\% & 15\% \\
\cmidrule(lr){2-6}
& Meta & LLaMA 3.1 405B Instruct & 9\% & 97\% & 13\% \\
&  & LLaMA 3.1 70B Instruct & 3\% & 97\% & 7.7\% \\
&  & LLaMA 4 Maverick 17B Instruct & 8\% & 94\% & 13\% \\
&  & LLaMA 4 Scout 17B Instruct & 2\% & 76\% & 2.4\% \\
\cmidrule(lr){2-6}
& Alibaba & Qwen2.5 72B Instruct Turbo & 0\% & 94\% & 2.7\% \\
\bottomrule
\end{tabular}
}
\end{table}

%% file: tables/success_k.tex
\begin{table}[h]
\centering
\caption{Model Performance with Multiple Independent Trials}
\label{tab:pass-at-k-metrics}
\resizebox{\textwidth}{!}{
\begin{tabular}{lcccc}
\toprule
\textbf{Model} & \textbf{Average Correctness} & \textbf{pass@1} & \textbf{pass@2} & \textbf{pass@5} \\
\midrule
Claude 3.7 Sonnet & \textbf{28.8\%} & \textbf{34\%} & 36\% & 41\% \\
Gemini 2.5 Pro (03-25 Preview) & 26.4\% & 28\% & \textbf{38\%} & \textbf{47\%} \\
DeepSeek R1 & 24.6\% & 24\% & 33\% & 46\% \\
OpenAI o3 & 22.8\% & 23\% & 30\% & 45\% \\
LLaMA 3.1 405B Instruct & 6.8\% & 8\% & 11\% & 13\% \\
\bottomrule

\end{tabular}
}
\begin{flushleft}
\small
\textbf{Average Correctness}: Average success rate across all trials (all tests passed) \\
\textbf{pass@1}: Probability of solving the task in a single attempt \\
\textbf{pass@2}: Probability of solving the task in at least one of two attempts \\
\textbf{pass@5}: Probability of solving the task in at least one of five attempts \\
\end{flushleft}
\end{table}

%% file: figures/error-distribution.tex



\begin{figure}[ht]
    \centering
    \makebox[\textwidth][c]{%
        \hspace{-0.1\textwidth}%
        \includegraphics[width=0.850\textwidth]{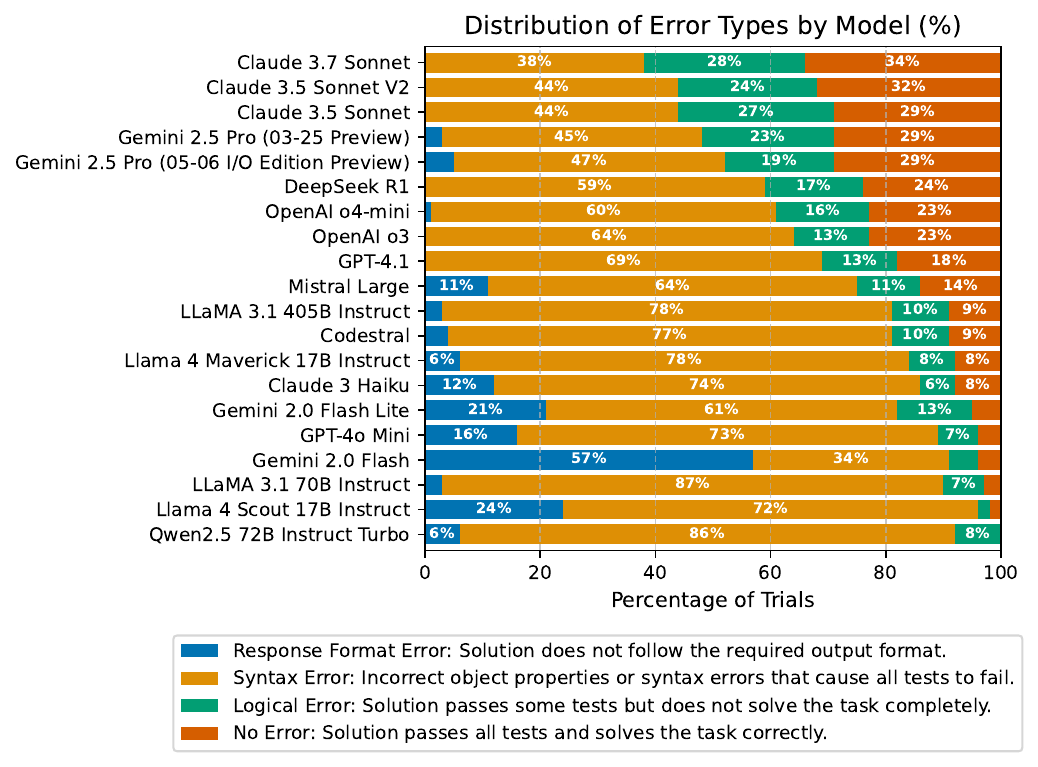}%
    }
    \caption{
        Distribution of error types across trials. \textit{Syntax errors} are the dominant failure type (40–85\% of errors), primarily stemming from incorrect property usage in CDK constructs. \textit{Logical errors} (5–30\%) occur when models produce syntactically valid code that fails to meet task requirements or capture context. \textit{Response format errors} are relatively rare, with Claude and OpenAI models demonstrating particular strength in format adherence.
    }
    \label{fig:error-dist}
\end{figure}

%% file: tables/two-turn-general-improvement-and-rag.tex
\begin{table}
\centering
\caption{Comparison of One-Turn vs. Two-Turn LLM Performance (with and without RAG).}
\label{tab:two-turn-comparison}
\resizebox{\textwidth}{!}{
\begin{tabular}{l|c|cc|cc|c|c|cc|cc|c}
\toprule
\multirow{3}{*}{\textbf{Model}} & \multicolumn{6}{c|}{\textbf{Correctness}~($\uparrow$)} & \multicolumn{6}{c}{\textbf{Passed Tests Share}~($\uparrow$)} \\
\cmidrule(lr){2-7} \cmidrule(lr){8-13}
 & \multirow{2}{*}{One-Turn} & \multicolumn{2}{c|}{Two-Turn} & \multicolumn{2}{c|}{Two-Turn + RAG} & \multirow{2}{*}{Best} & \multirow{2}{*}{One-Turn} & \multicolumn{2}{c|}{Two-Turn} & \multicolumn{2}{c|}{Two-Turn + RAG} & \multirow{2}{*}{Best} \\
 & & $V_L$ & $V_H$ & $V_L$ & $V_H$ & & & $V_L$ & $V_H$ & $V_L$ & $V_H$ & \\
\midrule
Claude 3.7 Sonnet & 34.0\% & 44.0\% & 64.0\% & 51.0\% & 60.0\% & \textbf{64.0\%} & 53.1\% & 49.5\% & 71.6\% & 60.6\% & 66.1\% & \textbf{71.6\%} \\
Claude 3.5 Sonnet V2 & 32.0\% & 52.0\% & 56.0\% & 52.0\% & 65.0\% & \textbf{65.0\%} & 47.0\% & 60.9\% & 66.1\% & 65.5\% & 74.2\% & \textbf{74.2\%} \\
Gemini 2.5 Pro (03-25 Preview) & 29.0\% & 41.0\% & 40.0\% & 41.0\% & 33.0\% & \textbf{41.0\%} & 42.5\% & 41.5\% & 41.4\% & 41.3\% & 33.0\% & \textbf{42.5\%} \\
DeepSeek R1 & 24.0\% & 40.0\% & 43.0\% & 45.0\% & 39.0\% & \textbf{45.0\%} & 34.9\% & 49.0\% & 51.7\% & 51.9\% & 49.5\% & \textbf{51.9\%} \\
GPT-4.1 (2025-04-14) & 18.0\% & 40.0\% & 48.0\% & 46.0\% & 26.0\% & \textbf{48.0\%} & 26.4\% & 48.0\% & 55.9\% & 56.7\% & 30.9\% & \textbf{55.9\%} \\
Mistral Large & 14.0\% & 21.0\% & 23.0\% & 14.0\% & 17.0\% & \textbf{23.0\%} & 20.3\% & 24.2\% & 28.6\% & 17.9\% & 23.7\% & \textbf{28.6\%} \\
LLaMA 4 Maverick 17B Instruct & 8.0\% & 15.0\% & 18.0\% & 21.0\% & 21.0\% & \textbf{21.0\%} & 13.0\% & 20.5\% & 23.7\% & 25.6\% & 27.3\% & \textbf{27.3\%} \\
\bottomrule
\end{tabular}
}
\begin{flushleft}
\small
\textbf{Correctness}: Percentage of completions where the solution passed all test cases. \\
\textbf{Passed Tests Share}: Average percentage of unit tests passed, including partial completions. \\
\textbf{Two-Turn + RAG}: Two-step prompting approach with retrieval-augmented generation support. \\
\textbf{$V_L$/$V_H$}: Low/High verbosity configurations.
\end{flushleft}
\end{table}

%% file: section/discussion.tex
\section{Conclusion}
\label{sec:discussion}

This work introduces \benchmark{}, the first benchmark designed to evaluate language models in modifying imperative infrastructure-as-code frameworks, specifically AWS CDK, on realistic cloud development tasks. \benchmark{} presents challenging IaC instances, each requiring LLMs to understand modification instructions, reason about cloud resource dependencies, and generate precise code changes within existing CDK repositories, verified automatically via unit tests. Our analysis reveals that these tasks, mirroring iterative enterprise workflows, pose significant difficulties for current models; our extensive experiments show that the best performing LLM Claude 3.7 struggles, with the top single-attempt success rate reaching only 34\%, although multi-turn agentic approaches show promise, boosting performance up to 65\%. Other strong LLMs, including reasoning models such as DeepSeek r1, falls short with only 24\% pass rate. 

\benchmark{} currently centers on AWS CDK (Python), which is a popular way of building IaC applications in the cloud. We hope in the future to broaden the benchmark to encompass additional cloud services and IaC frameworks. Similarly, while not a focus of the present work, a natural extension is to enable agents with various tools such as search engine, bash script, python environment and others that could facilitate development of IaC code modifications. Our results with two-turn LLMs show that this is a promising direction. Finally, although we employ execution-based testing for model evaluation, this method alone cannot fully assess the quality of generated infrastructure code. Automatically generated solutions may be syntactically correct and pass unit tests, yet still suffer from inefficiencies, misconfigurations, or security risks when deployed. Future evaluation strategies could incorporate static analysis tools, cost simulation, or LLM-as-a-Judge review to address these gaps.

%% file: appx/statistics.tex
\section{\benchmark{} Characteristics}
\label{sec:benchmark:features-of-benchmark}

\begin{figure}[H]
\centering
\includegraphics[width=0.85\textwidth]{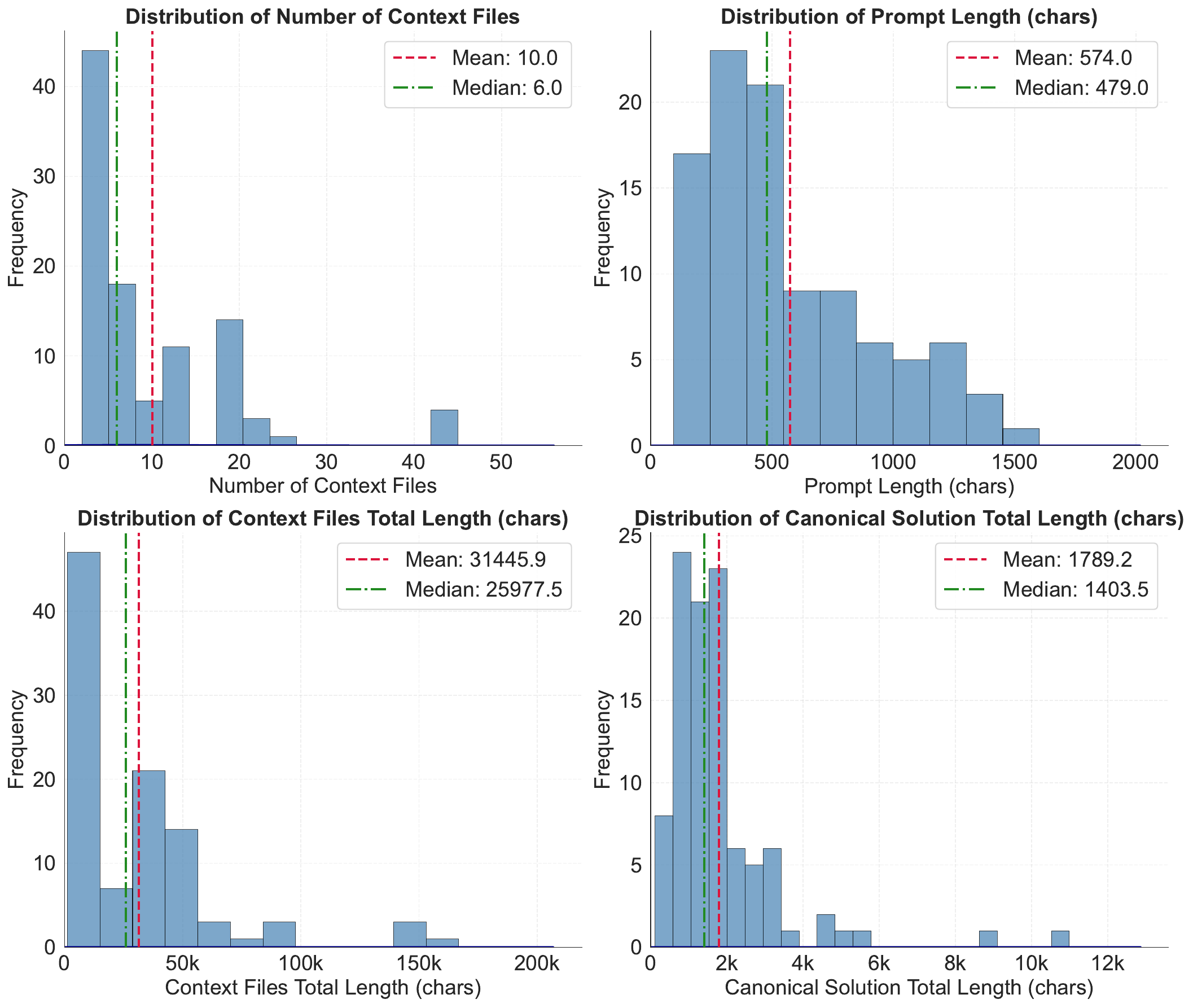}
\caption{\benchmark{} Statistics on Context Size and Solution Length}
\label{fig:dataset-stats}
\end{figure}

\benchmark{} tasks contain examples for multiple CDK versions ranging from 2.0.0 to more recent version like 2.189.1, with the majority of tasks suitable for 2.178.2 and newer (see Figure~\ref{fig:cdk-versions}). The tasks also vary in number and length of context files, prompt length and canonical solution length. The key dataset statistics are illustarted in Figure~\ref{fig:dataset-stats}.

%% file: appx/cdk_versions.tex
\section{CDK Versions Distribution}
\label{app:cdk_versions}


\begin{figure}[H]
\centering
\includegraphics[width=0.65\textwidth]{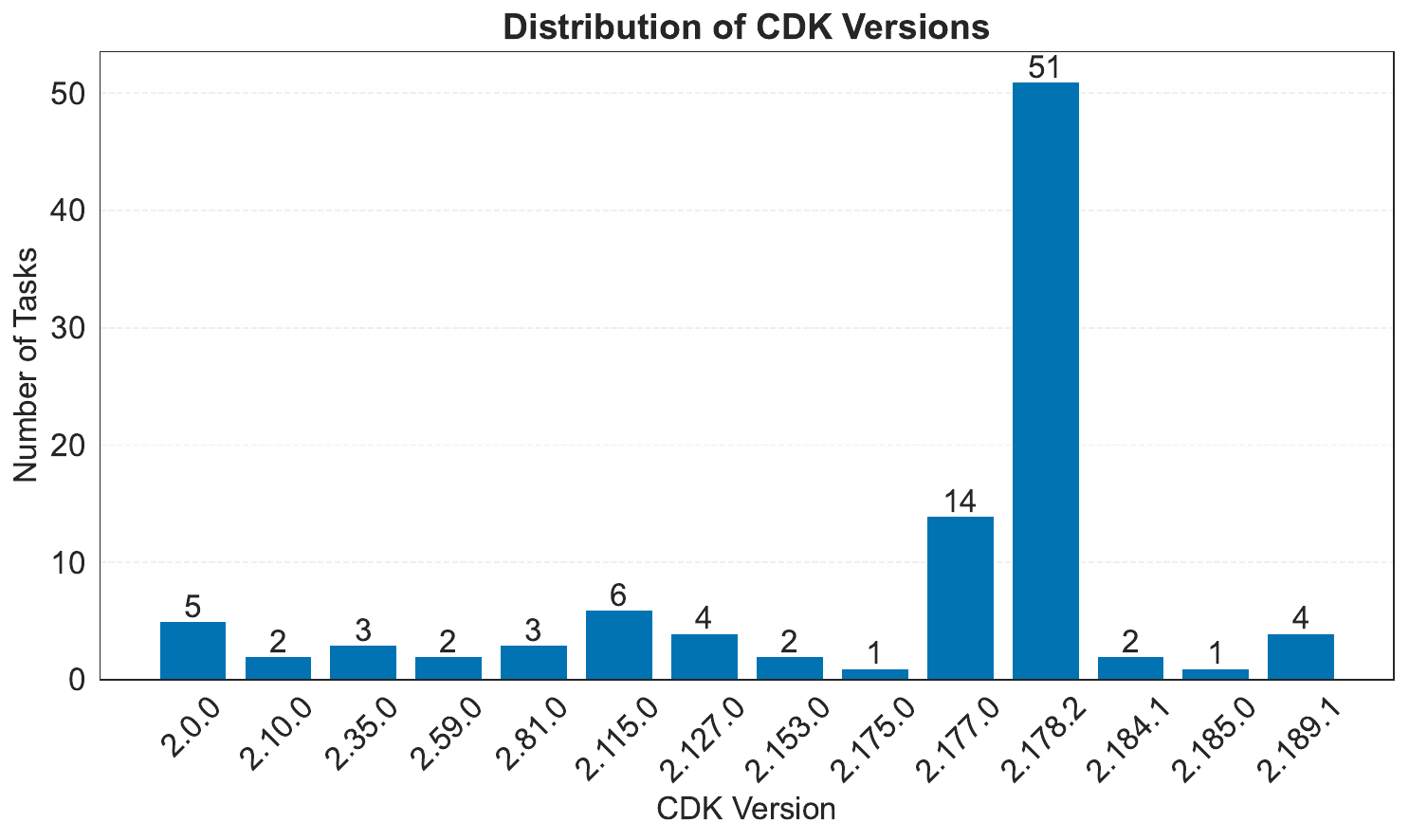}
\caption{Distribution of AWS CDK Versions Across \benchmark{} Tasks}
\label{fig:cdk-versions}
\end{figure}

Figure~ \ref{fig:cdk-versions} shows the distribution of AWS CDK versions across the SWE-InfraBench tasks. The benchmark predominantly utilizes version 2.178.2, highlighting a concentration on recent library versions, while also including several earlier versions to ensure diversity and represent real-world variability in CDK projects.

%% file: appx/source-licences.tex
\section{Source Repositories Licences}
\label{app:source-licences}

\benchmark{} tasks are derived from open-source repositories and custom-developed sources. Note that original code snippets were substantially modified to create the tasks. See Table~\ref{tab:os-repos} for details on open-source repositories.

\begin{table}[H]
\centering
\caption{Open-source repositories used as sources for \benchmark{} tasks}
\label{tab:os-repos}
\resizebox{\textwidth}{!}{
\begin{tabular}{lll}
\toprule
\textbf{Repository Name} & \textbf{Source} & \textbf{License} \\
\midrule
aws-cdk-examples & aws-samples/aws-cdk-examples & Apache-2.0 \\
generative-ai-cdk-constructs-samples & aws-samples/generative-ai-cdk-constructs-samples & Apache-2.0 \\
generative-ai-ml-latam-samples & aws-samples/generative-ai-ml-latam-samples & MIT-0 \\
aws-cdk-lambda-import-export-redshift-ddl & aws-samples/aws-cdk-lambda-import-export-redshift-ddl & MIT-0 \\
amazon-elasticache-demo-using-aws-cdk & aws-samples/amazon-elasticache-demo-using-aws-cdk & MIT-0 \\
deploy-langfuse-on-ecs-with-fargate & aws-samples/deploy-langfuse-on-ecs-with-fargate & MIT-0 \\
\bottomrule
\end{tabular}
}
\begin{flushleft}
\small
All repository URLs are prefixed with \url{https://github.com/}
\end{flushleft}
\end{table}

%% file: appx/tasks_teaser.tex
\section{Tasks Preview}
\label{section:tasks-teaser}

\subsection{Prompt Examples}
\label{subsection:tasks-prompt-examples}

\benchmark{} includes a diverse set of prompts that test various aspects of infrastructure-as-code generation. See examples of such prompts below:

\begin{tcolorbox}[
    enhanced,
    colback=blue!5!white,
    colframe=blue!75!black,
    title=API Gateway Integration,
    fonttitle=\bfseries\sffamily,
    boxrule=0.5pt,
    arc=2mm,
    left=5mm,
    right=5mm
]
Generate code to create an API Gateway SampleAPI-EventBridge-Multi-Consumer that integrates with the event producer with proxy integration. Do not add a catch-all route in api routing (use proxy=False). Add a resource named 'items' to the root of the API and create a POST method for this resource.
\end{tcolorbox}

\begin{tcolorbox}[
    enhanced,
    colback=green!5!white,
    colframe=green!75!black,
    title=Bedrock Agent Setup,
    fonttitle=\bfseries\sffamily,
    boxrule=0.5pt,
    arc=2mm,
    left=5mm,
    right=5mm
]
Create a Bedrock agent using CDK with the following requirements in the CfnAgent construct.

Set the agent name, description, model and instruction from the class attributes.

Also, set the agent resource role ARN from the previously created role and add idle session time of 600 seconds.

We want the draft version to be always in sync via auto preparation and add test alias tags.

Finally include action groups and collaboration if provided and ensure the removal policy destroys the agent when the stack is deleted
\end{tcolorbox}

\begin{tcolorbox}[
    enhanced,
    colback=orange!5!white,
    colframe=orange!75!black,
    title=WhatsApp IAM Policy,
    fonttitle=\bfseries\sffamily,
    boxrule=0.5pt,
    arc=2mm,
    left=5mm,
    right=5mm
]
Create CDK code to add IAM policies to a Lambda function, granting permissions for social-messaging, transcribe, and bedrock services.

The policies should have separate policies for

1) allow sending whatsapp messages and getting media from them for all cell numbers in all regions and accounts

2) access to any transcribe action on all resources

3) and access to invoke* all agents, inference profiles and models in oregon
\end{tcolorbox}

\subsection{Task Example}
\label{subsection:tasks-example-teaser}
Figure~\ref{fig:task-example} illustrates the structure of a task from \benchmark{}. Each task is stored as a JSON file containing the prompt, context files, canonical solution, and tests.

\begin{figure}[H]
\begin{tcolorbox}[
    enhanced,
    colback=gray!5,
    colframe=gray!75!black,
    title=Example Task: API Gateway Integration with EventBridge,
    fonttitle=\bfseries,
    boxrule=0.5pt,
    arc=2mm,
    left=3mm,
    right=3mm
]
\begin{lstlisting}[
    frame=none,
    basicstyle=\ttfamily\footnotesize,
    breaklines=true
]
{
  "task_id": "67ad6ef1-fb3e-45e6-b5e1-83ae385528b5",
  "entry_point": "api-eventbridge-lambda+api_gateway_integration",
  "prompt": "Generate code to create an API Gateway SampleAPI-EventBridge-Multi-Consumer 
            that integrates with the event producer with proxy integration. 
            Do not add a catch-all route in api routing (use proxy=False). 
            Add a resource named 'items' to the root of the API and create a 
            POST method for this resource.",
  "cdk_version": "2.178.2",
  "context": {
    "app.py": "#!/usr/bin/env python3\n\nfrom aws_cdk import App\n\nfrom api_eventbridge_lambda.api_eventbridge_lambda import ApiEventBridgeLambdaStack...",
    "api_eventbridge_lambda/api_eventbridge_lambda.py": "from constructs import Construct\nfrom aws_cdk import...",
    "lambda/event_consumer_lambda.py": "import json\nimport logging\n\nlogger = logging.getLogger()...",
    "lambda/event_producer_lambda.py": "import json\nimport boto3\nimport datetime..."
  },
  "canonical_solution": {
    "api_eventbridge_lambda/api_eventbridge_lambda.py": [
      "--- without_solution\n\n+++ with_solution\n\n@@ -102,0 +103,7 @@\n\n+        # defines an API Gateway REST API resource backed by our \"atm_producer_lambda\" function.\n\n+        api = api_gw.LambdaRestApi(self, 'SampleAPI-EventBridge-Multi-Consumer',\n\n+                             handler=event_producer_lambda,\n\n+                             proxy=False\n\n+                             )\n\n+        items = api.root.add_resource(\"items\")\n\n+        items.add_method(\"POST\")  # POST /items\n"
    ]
  },
  "tests": {
    "test_api_gateway_integration.py": "import aws_cdk as cdk\nfrom aws_cdk.assertions import Template, Match..."
  }
}
\end{lstlisting}
\end{tcolorbox}
\caption{Example task from InfraBench showing the JSON structure. Context and test files are truncated for brevity.}
\label{fig:task-example}
\end{figure}

%% file: appx/prompts.tex
\section{Prompt Templates}

\subsection{Critic Prompt Template}
\label{app:critic_prompt}

The following is the template used for the critic in the InfraBench dataset construction process:

\begin{tcolorbox}[breakable, title=Critic Prompt Template]
As an AI assistant specialized in Infrastructure as Code, your task is to critique a dataset item created for an LLM code generation challenge.

You will be given:

1. A repository description

2. The full repository code

3. A pre-suggestion with a specific code section to mask

4. The actual code that was masked

5. The generated human language prompt

6. The generated test file

Your job is to critically evaluate:

1. Whether the prompt accurately describes what needs to be implemented

2. Whether the tests effectively validate all requirements stated in the prompt

3. Whether the prompt and tests are general enough to allow various valid solutions that preserve the functionality of the masked code while remaining compatible with the overall repository structure

\# REPOSITORY DESCRIPTION
\begin{lstlisting}[breaklines=true]
[Repository description is provided here]
\end{lstlisting}

\# REPOSITORY CONTENT
\begin{lstlisting}[breaklines=true]
[Repository files are provided here]
\end{lstlisting}

\# PRE-SUGGESTION
\begin{lstlisting}[breaklines=true]
Item Name: [Item name]
File Path: [File path]
Start Line: [Start line]
End Line: [End line]
Complexity: [Complexity]
Rationale: [Rationale]
\end{lstlisting}

\# ACTUAL CODE TO BE MASKED
\begin{lstlisting}[breaklines=true]
[Masked code is provided here]
\end{lstlisting}

\# GENERATED PROMPT
\begin{lstlisting}[breaklines=true]
[Generated prompt is provided here]
\end{lstlisting}

\# GENERATED TEST FILE

Path: [Test file path]
\begin{lstlisting}[breaklines=true]
[Test file content is provided here]
\end{lstlisting}

\# GENERATOR NOTES
\begin{lstlisting}[breaklines=true]
[Generator notes are provided here]
\end{lstlisting}

\# CRITIQUE GUIDELINES

\#\# Prompt vs Functional Requirements Evaluation

1. Does the prompt clearly describe ALL the necessary functional aspects needed for the masked code?

2. Is it sufficiently detailed for someone to implement the solution correctly without seeing the masked code?

3. Are there any ambiguities or missing requirements that would prevent a correct implementation?

4. Does it avoid revealing the actual implementation details while still being complete?

5. If something can be inferred unequivocally from the repository code, it does not need to be specified in the prompt.

\#\# Generality of Tests Evaluation

1. Tests should be general enough so that they pass if a developer or LLM follows the prompt accurately, regardless of the specific implementation details.

2. The prompt should give only the minimum necessary instructions needed to explain the functional requirements, while considering how the tests are built.

3. Tests should verify functionality rather than specific implementation approaches - they should allow for multiple valid solution patterns that fulfill the prompt.

4. Evaluate if tests are overly restrictive by enforcing a particular implementation approach when other valid approaches could fulfill the same requirements.

5. Tests won't be accepted if they don't pass with the original masked code (tested elsewhere) but consider as well if other reasonable implementations would pass.

\#\# Test Evaluation

For each test in the test file:

1. What is this test specifically checking for?

2. Is this test testing for something that's explicitly stated in the prompt?

3. A test is valid if it tests integration with functionality that exists elsewhere in the code base (not just the masked section).

4. Is the test appropriately written to verify the requirement?

5. If the test might fail with some valid implementations (including the original masked code), should the prompt be more explicit or should the test be less restrictive?

\#\# Test Completeness Evaluation

1. Do the tests collectively verify ALL requirements mentioned in the prompt?

2. Are there any requirements in the prompt that aren't tested?

3. Are there any tests for requirements not mentioned in the prompt?

4. Are all edge cases and error conditions properly tested?

Return the response in this JSON format:
\begin{lstlisting}[breaklines=true]
{
    "prompt_vs_functional": {
        "explanation": "Detailed explanation of whether the prompt accurately describes all necessary functional aspects of the masked code",
        "corrections": "Specific corrections with concrete implementation suggestions - provide exact wording changes or additions to the prompt", // Only include if there are issues
        "example_improvements": "Suggested rewrites of problematic sections with specific language", // Only include if there are issues
        "accept": true|false  // Conclusion based on the explanation
    },
    "generality": {
        "explanation": "Analysis of whether the tests allow for multiple valid implementations that satisfy the requirements in the prompt",
        "issues": "Identification of any tests that could fail with valid implementations that follow the prompt", // Only include if there are issues
        "suggested_improvements": "Concrete suggestions for making tests more general while still validating functionality", // Only include if needed
        "accept": true|false  // Whether the tests are sufficiently general
    },
    "tests": {
        "test_name_1": {
            "purpose": "What this test is checking for",
            "coverage": "Explanation of how this relates to prompt requirements",
            "suggested_improvements": {
                "explanation": "Why the test needs improvement", // Only include if needed
                "code_snippet": "Complete improved version of the test with code fixes", // Provide actual code implementation
                "rationale": "Explanation of why this implementation is better"
            }, // Only include if needed
            "accept": true|false  // Conclusion based on the critic's analysis of the test quality
        },
        "test_name_2": {
            "purpose": "What this test is checking for",
            "coverage": "Explanation of how this relates to prompt requirements",
            "suggested_improvements": {
                "explanation": "Why the test needs improvement", // Only include if needed
                "code_snippet": "Complete improved version of the test with code fixes", // Provide actual code implementation
                "rationale": "Explanation of why this implementation is better"
            }, // Only include if needed
            "accept": true|false  // Conclusion based on the critic's analysis of the test quality
        }
        // Add entries for each test in the test file
    },
    "tests_completeness": {
        "explanation": "Analysis of whether the tests completely cover all prompt requirements",
        "missing_tests": ["list", "of", "requirements", "that", "should", "be", "tested", "but", "aren't"],
        "corrections": "Specific additional tests needed if the evaluation fails", // Only include if there are issues
        "accept": true|false  // Conclusion based on the analysis
    },
    "feedback": "Detailed feedback explaining all issues and providing clear guidance for improvements"
}
\end{lstlisting}

Be rigorous in your evaluation. The goal is to ensure high-quality dataset items that will effectively test LLM code generation capabilities.
Only provide the JSON response, no additional explanation. The response will be parsed with json.loads(response) so be sure json format is correct. Instead of triple-quotes use \textbackslash n characters.

\# IMPORTANT CUSTOM INSTRUCTIONS
\begin{lstlisting}[breaklines=true]
[Custom instructions from human reviewers are provided here when available. These instructions provide special guidance in the evaluation process.]
\end{lstlisting}
\end{tcolorbox}

\subsection{Generator Prompt Template}
\label{app:generator_prompt}
The following is the template used for the generator in the InfraBench dataset construction process:

\begin{tcolorbox}[breakable, enhanced, title=Generator Prompt Template]
As an AI assistant specialized in Infrastructure as Code, your task is to create a high-quality dataset item for an LLM code generation challenge.

You will be given:

1. A repository description

2. A pre-suggestion with a specific code section to mask

3. The full repository content

Based on this information, you need to:

1. Create a clear, detailed human language prompt describing what code needs to be generated

2. Develop comprehensive pytest tests that validate the generated code meets all requirements

\# REPOSITORY DESCRIPTION
\begin{lstlisting}[breaklines=true]
[Repository description is provided here]
\end{lstlisting}

\# PRE-SUGGESTION
\begin{lstlisting}[breaklines=true]
Item Name: [Item name]
File Path: [File path]
Start Line: [Start line]
End Line: [End line]
Complexity: [Complexity]
Rationale: [Rationale]
\end{lstlisting}

\# PREVIOUS FEEDBACK

When available, this section may include:

\#\# PREVIOUS TASK SUGGESTION
\begin{lstlisting}[breaklines=true]
{
  "item_name": "example_item",
  "file_path": "path/to/file",
  "insert_points": {
    "start_line": 10,
    "end_line": 20
  },
  "prompt": "previous prompt text",
  "test_file": {
    "path": "tests/test_example.py",
    "content": "previous test file content"
  },
  "generator_notes": "previous notes"
}
\end{lstlisting}

\#\# IF THERE ARE VALIDATION ERRORS

The tests failed when run against the original masked code. Here are the errors:
\begin{lstlisting}[breaklines=true]
Error details from validation step
\end{lstlisting}
Make sure the original code will pass the tests, and write them correctly according to what you see in these error logs.

\#\# CRITIQUE FEEDBACK
\begin{lstlisting}[breaklines=true]
[Detailed feedback from the critique step explaining issues with the previous suggestion.]
\end{lstlisting}

\# GUIDELINES FOR CREATING A QUALITY DATASET ITEM

\#\# For the human language prompt:

1. COMPLETENESS: Describe ALL functional aspects needed for the code.

2. GENERALITY: Allow for multiple possible valid solutions that fulfill the requirements.

3. CLARITY: Be specific about requirements but avoid dictating implementation specifics.

4. PRECISION: Include all requirements that would allow someone to implement the solution correctly.

5. CONTEXT: Describe the functionality, purpose, and integration with other components.

6. ESSENTIAL DETAILS ONLY: Specify necessary parameters and behaviors, but avoid over-constraining the solution.

7. NO SPOILERS: DO NOT include the actual implementation details or code snippets.

8. IMPLEMENTATION FREEDOM: Focus on "what" needs to be achieved, not "how" it must be done.

9. NO HINTS: If something can be inferred unequivocally from the repository code, it does not need to be specified in the prompt.

\#\# For the test file:
1. COMPREHENSIVE COVERAGE: Tests must verify ALL aspects mentioned in the prompt

2. EXPLICIT PURPOSE: Each test should clearly indicate what requirement it's checking

3. APPROPRIATE VERIFICATION: Tests must use assertions that correctly validate the implementation

4. PRACTICALITY: Tests MUST pass when run against the original code that will be masked

5. ROBUSTNESS: Tests should fail if important requirements are not met

6. COMPLETENESS: No requirement from the prompt should be left untested

7. STRUCTURE: Use appropriate fixtures and mocks, follow pytest best practices

\# REPOSITORY CONTENT
File to be masked: [File path]
\begin{lstlisting}[breaklines=true]
[File content is provided here]
\end{lstlisting}

\begin{lstlisting}[breaklines=true]
[All other relevant context files are provided here]
\end{lstlisting}

Return the response in this JSON format:
\begin{lstlisting}[breaklines=true]
{
    "item_name": "SAME_AS_PRE_SUGGESTION",  
    "file_path": "path/to/file",  
    "insert_points": {
        "start_line": number,  
        "end_line": number     
    },
    "prompt": "detailed description of what needs to be generated",
    "test_file": {
        "path": "tests/test_something.py",
        "content": "complete content of the test file including imports, fixtures, and test cases"
    },
    "generator_notes": "You don't need to accept all the suggestions from the feedback, but give an explanation of your approach, learnings regarding test syntax for the cannonical solution to pass, detailed design decisions, rationale for implementation choices, and how you've ensured generality in the prompt and tests. There will be a new critic, so explain your decisions without assuming the critic understands the current feedback. IMPORTANT: give an analysis of possible flaws in the generated prompt and tests focusing on its generality (if various valid solutions would be accepted), coverage and alignment between prompt and tests, etc."
}
\end{lstlisting}
Ensure that:

1. The insert\_points are within the pre-suggestion range

2. The prompt is comprehensive and covers ALL requirements

3. The test\_file content is complete and will validate ALL requirements

4. The tests MUST pass when run against the original masked code

Only provide the JSON response, no additional explanation. 

\# IMPORTANT CUSTOM INSTRUCTIONS
\begin{lstlisting}[breaklines=true]
[Custom instructions from human reviewers are provided here when available. These instructions provide special guidance in the generation process.]
\end{lstlisting}
\end{tcolorbox}

\subsection{Zero-shot Direct Solver}
\label{subsection:direct-solver-prompts}

\begin{tcolorbox}[title=\textbf{Anthropic template}, enhanced, breakable, colback=gray!5, colframe=gray!50!black, fonttitle=\bfseries]
You are tasked with implementing a solution based on the following prompt:

\begin{verbatim}
<example_prompt>
{example_prompt}
</example_prompt>
\end{verbatim}

Ensure the solution is compatible with AWS Python CDK version aws-cdk-lib = \verb+{cdk_version}+.

You have access to the following context files:
\begin{verbatim}
<context_files>
{context_files}
</context_files>
\end{verbatim}

Your task is to provide git-style unified diffs that show the changes needed to implement the solution.
For each file that needs changes, provide a unified diff.
Only add content, do not remove any lines

Provide your response as a JSON object where:
- Keys are the file paths
- Values are arrays containing the unified diffs for each change section

The diff format should be:
\begin{verbatim}
<output_format>
--- without_solution
+++ with_solution
@@ -line,count +line,count @@
    context lines
+added lines
    context lines
</output_format>
\end{verbatim}

Example response format:
\begin{verbatim}
<example_response>
{
    "path/to/file1.py": [
        "--- without_solution\\n+++ with_solution\\n
        @@ -10,3 +10,5
        @@\\n     existing_line\+    
        new_line1\\n+    new_line2\\n
        existing_line"
    ]
}
</example_response>
\end{verbatim}

Only provide the JSON response, no additional explanation. The response will be parsed with \verb+json.loads(response_text)+ so make sure the string is correct JSON. Do NOT include \verb+```json+

Ensure the diffs include proper line numbers and context.
\end{tcolorbox}

\begin{tcolorbox}[title=\textbf{Default template}, enhanced, breakable, colback=gray!5, colframe=gray!50!black, fonttitle=\bfseries]
You are tasked with implementing a solution based on the following prompt:

\begin{verbatim}
{example_prompt}
\end{verbatim}

Ensure the solution is compatible with AWS Python CDK version aws-cdk-lib = \verb+{cdk_version}+.

You have access to the following context files:
\begin{verbatim}
{context_files}
\end{verbatim}

Your task is to provide git-style unified diffs that show the changes needed to implement the solution.
For each file that needs changes, provide a unified diff.
Only add content, do not remove any lines

Provide your response as a JSON object where:
- Keys are the file paths
- Values are arrays containing the unified diffs for each change section

The diff format should be:
\begin{verbatim}
--- without_solution
+++ with_solution
@@ -line,count +line,count @@
    context lines
+added lines
    context lines
\end{verbatim}

Example response format:
\begin{verbatim}
{
    "path/to/file1.py": [
        "--- without_solution\\n+++ with_solution\\n
        @@ -10,3 +10,5
        @@\\n     existing_line\+    new_line1
        \\n+    new_line2\\n
        existing_line"
    ]
}
\end{verbatim}

Only provide the JSON response, no additional explanation. The response will be parsed with \verb+json.loads(response_text)+ so make sure the string is correct JSON. Do NOT include \verb+```json+

Ensure the diffs include proper line numbers and context.
\end{tcolorbox}

\subsection{Zero-shot Two-Turn Solver}

\begin{tcolorbox}[title=\textbf{First Turn Prompt}, enhanced, breakable, colback=gray!5, colframe=gray!50!black, fonttitle=\bfseries, width=\linewidth]
\begin{verbatim}
# Task overview
You are tasked with implementing a solution 
based on the following prompt:
{example_prompt}

# Steps
- Have a look at the version of aws-cdk-lib
    - Current version is {cdk_version}
- Read each all of the context files ("# Context files" section)
- Try to understand what should be done
- Read task overview 
("# Task overview" section)
- If provided, read related AWS documentation 
("# Related documentation" section)
- Provide solution with a specified format 
("# Response format" section)

# Response format
## General guidelines
Provide your solution as a JSON object with:
1. File paths as keys
2. Arrays of unified diffs as values

## Diff format specification
Each diff must follow this structure:
```
--- without_solution
+++ with_solution
@@ -line,count +line,count @@
 context lines
+added lines
 context lines
```

## Important rules:
- Only ADD content, do not remove any lines
- Include proper line numbers and context
- Ensure diffs are properly formatted

## Example JSON response:
```json
{{
    "path/to/file1.py": [
        "--- without_solution\n+++ with_solution\n
        @@ -10,3 +10,5 @@\n
        existing_line\n+    new_line1\n+    new_line2\n
        existing_line"
    ]
}}
```

## Additional guidelines
- Provide ONLY the JSON response
- No additional explanation
- Response must be valid JSON (will be parsed with json.loads())
- If you want to think before returning response, 
be short and concise

# Context files
Here are the context files to analyze:
{context_files}
\end{verbatim}
\end{tcolorbox}

\begin{tcolorbox}[title=\textbf{Second Turn Prompt}, enhanced, breakable, colback=gray!5, colframe=gray!50!black, fonttitle=\bfseries, width=\linewidth]
\begin{verbatim}
# Task overview
You are tasked with fixing errors in an LLM-generated solution
based on the following initial prompt:
{example_prompt}

# Steps
- Have a look at the version of aws-cdk-lib
    - Current version is {cdk_version}
- Read each all of the context files ("# Context files" section)
- Try to understand what should be done
- Read task overview ("# Task overview" section)
    - This is an initial formulation of the task 
    that LLM have used to generate its solution
- Read previous attempt solution code 
("# Previous attempt" section)
    - This is the solution generated by LLM which fails checks
- Read error traceback ("# Error message" section)
    - Interpret error message
    - Identify specific error type (syntax, type, logic, etc.)
- If provided, read related AWS documentation 
("# Related documentation" section)
- Provide error analysis
    - Put it in "error_analysis"
- Provide a working solution with a specified format
    - Put it in "regenerated_solution"

# Response format
## General guidelines
Provide your solution as a JSON object with:
1. "error_analysis": Your complete error analysis
2. "regenerated_solution": Object containing file paths and 
their diffs

## Diff format specification
Each diff must follow this structure:
```
--- without_solution
+++ with_solution
@@ -line,count +line,count @@
 context lines
+added lines
 context lines
```

## Important rules:
- Only ADD content, do not remove any lines
- Include proper line numbers and context
- Ensure diffs are properly formatted

## Example JSON response:
```json
{{
    "error_analysis": "The error occurred because...",
    "regenerated_solution": {{
        "path/to/file1.py": [
            "--- without_solution\n+++ with_solution\n
            @@ -10,3 +10,5
            @@\n     existing_line\n+    new_line1\n
            +    new_line2\n
            existing_line"
        ]
    }}
}}
```

## Additional guidelines
- Provide ONLY the JSON response
- No additional explanation
- Response must be valid JSON (will be parsed with json.loads())
- If you want to think before returning response, 
be short and concise

# Related documentation
Here are supporting documentation pieces:
```markdown
{documentation}
```

# Context files
Here are the context files to analyze:
{context_files}

# Previous attempt
Previous solution, which needs fixing:
{solution_code}

# Error message
Error message of the previous solution:
{error_message}

\end{verbatim}
\end{tcolorbox}

\subsection{Zero-shot Two-Turn Solver (RAG)}

For this configuration, same prompt template from previous configuration is used for second turn. First turn template is different from the default two-turn solver configuration and allows in the same time to generate solution and keywords to search for in the documentation:

\begin{tcolorbox}[title=\textbf{First Turn Prompt}, enhanced, breakable, colback=gray!5, colframe=gray!50!black, fonttitle=\bfseries, width=\linewidth]
\begin{verbatim}
# Task overview
You are tasked with implementing a solution 
based on the following prompt:
{example_prompt}

# Context files
Here are the context files to analyze:
{context_files}

# Steps
- Have a look at the version of aws-cdk-lib
    - Current version is {cdk_version}
- Read task overview ("# Task overview" section)
- Read each all of the context files ("# Context files" section)
- Try to understand what should be done
- If provided, read related AWS documentation 
("# Related documentation" section)
- Provide solution with a specified format 
("# Response format" section)
- Provide documentation support keywords

# Documentation Support
To confirm your implementation you should provide keywords in 
"search" 
in key:
- Include a "search" key in your JSON response
- Provide up to 5 specific keywords 
related to the AWS services
and features you need help with
- Example: "search": "data access policies opensearch"
    - Avoid using underscores or other similar 
    special characters, query should be human-readable
    - Keywords should reflect the resources used in the stack 
    or errors seen during execution, 
    i.e. 'opensearch' or 'appsync'
    - Avoid too generic keywords like 'aws' or 'cloudformation'
- Keywords you provide will be used in the next interaction
in case solution does not pass the tests

# Response format
## General guidelines
Provide your solution as a JSON object with:
1. File paths as keys
2. Arrays of unified diffs as values

## Diff format specification
Each diff must follow this structure:
```
--- without_solution
+++ with_solution
@@ -line,count +line,count @@
 context lines
+added lines
 context lines
```

## Important rules:
- Only ADD content, do not remove any lines
- Include proper line numbers and context
- Ensure diffs are properly formatted

## Example JSON response:
```json

{{
    "path/to/file1.py": [
        "--- without_solution\n+++ with_solution\n
        @@ -10,3 +10,5 
        @@\n     existing_line\n+    new_line1\n
        +    new_line2\n
        existing_line"
    ],
    "search": "..."
}}
```

## Additional guidelines
- Provide ONLY the JSON response
- No additional explanation
- Response must be valid JSON (will be parsed with json.loads())

------

Your JSON response (start with ```json):

\end{verbatim}
\end{tcolorbox}

%% file: appx/model_configs.tex
\section{Model Configurations}
\label{section:model-configs}

Table \ref{tab:model-configs} overviews the models used in our experiments and provides their invocation parameters. Some reasoning models (DeepSeek R1, OpenAI o3, OpenAI o4-mini) have \textbf{4 times bigger token budget}, Gemini 2.5 models require \textbf{5 times more}, reaching more than 20K maximum tokens per task. This was done to allow for longer outputs for reasoning models to avoid result truncation.

\begin{table}[H]
\centering
\caption{Models Invocation Parameters}
\label{tab:model-configs}
\resizebox{\textwidth}{!}{
\begin{tabular}{lll}
\toprule
{\bf Model} & {\bf Model ID} & {\bf Invocation Parameters} \\
\midrule
Claude 3 Haiku & anthropic.claude-3-haiku-20240307-v1:0 & max\_tokens=4096; temperature=0.25; top\_p=0.9 \\
\midrule
Claude 3.5 Sonnet & anthropic.claude-3-5-sonnet-20240620-v1:0 & max\_tokens=4096; temperature=0.25; top\_p=0.9 \\
\midrule
Claude 3.5 Sonnet V2 & anthropic.claude-3-5-sonnet-20241022-v2:0 & max\_tokens=4096; temperature=0.25; top\_p=0.9 \\
\midrule
Claude 3.7 Sonnet & claude-3-7-sonnet-latest & max\_tokens=4096; temperature=0.25 \\
\midrule
Codestral & codestral-latest & max\_tokens=4096; temperature=0.25 \\
\midrule
DeepSeek R1 & deepseek-ai/DeepSeek-R1 & max\_tokens=16384; temperature=0.25 \\
\midrule
GPT-4.1 & gpt-4.1 & max\_output\_tokens=4096; temperature=0.25 \\
\midrule
GPT-4o Mini & gpt-4o-mini & max\_output\_tokens=4096; temperature=0.25 \\
\midrule
Gemini 2.0 Flash & gemini-2.0-flash  & max\_output\_tokens=4096; temperature=0.25 \\
\midrule
Gemini 2.0 Flash Lite & gemini-2.0-flash-lite & max\_output\_tokens=4096; temperature=0.25 \\
\midrule
Gemini 2.5 Pro (03-25 Preview) & gemini-2.5-pro-preview-03-25 & max\_output\_tokens=20480; temperature=0.25 \\
\midrule
Gemini 2.5 Pro (05-06 I/O Edition Preview) & gemini-2.5-pro-preview-05-06 & max\_output\_tokens=20480; temperature=0.25 \\
\midrule
LLaMA 3.1 405B Instruct & meta.llama3-1-405b-instruct-v1:0 & max\_tokens=4096; temperature=0.25; top\_p=0.9 \\
\midrule
LLaMA 3.1 70B Instruct & us.meta.llama3-3-70b-instruct-v1:0 & max\_tokens=4096; temperature=0.25; top\_p=0.9 \\
\midrule
LLaMA 4 Maverick 17B Instruct & meta.llama4-maverick-17b-instruct-v1:0  & max\_tokens=4096; temperature=0.25 \\
\midrule
LLaMA 4 Scout 17B Instruct & meta.llama4-scout-17b-instruct-v1:0 & max\_tokens=4096; temperature=0.25 \\
\midrule
Mistral Large & mistral-large-latest  & max\_tokens=4096; temperature=0.25 \\
\midrule
OpenAI o3 & o3  & max\_output\_tokens=16384 \\
\midrule
OpenAI o4-mini & o4-mini  & max\_output\_tokens=16384 \\
\midrule
Qwen2.5 72B Instruct Turbo & Qwen/Qwen2.5-72B-Instruct-Turbo & max\_tokens=4096; temperature=0.25 \\
\bottomrule
\end{tabular}
}
\end{table}

%% file: appx/verbosity.tex
\section{Verbosity Configurations for Two-Turn Solvers}

Our experimental framework uses pytest as the testing library, with two distinct verbosity configurations designed to evaluate how different levels of feedback detail affect model performance. Here are the details on each of the configurations:

\input{tables/appx/verbosity-configurations}

%% file: tables/appx/verbosity-configurations.tex
\begin{table}[h]
\centering
\caption{Verbosity Parameters}
\label{tab:verbosity-configurations}
\begin{tabular}{p{0.25\linewidth}p{0.38\linewidth}p{0.38\linewidth}}
\toprule
\textbf{} & \textbf{Low-verbosity ($V_L$)} & \textbf{High-verbosity ($V_H$)} \\
\midrule
\multicolumn{3}{l}{\textbf{Configuration}} \\
Flags & \texttt{-q --tb=no --no-summary} & $\emptyset$ \\
\midrule
\multicolumn{3}{l}{\textbf{Output Features}} \\
Pass/fail counter & \checkmark & \checkmark \\
Exception & \checkmark & \checkmark \\
Traceback & $\times$ & \checkmark \\
Test files names & $\times$ & \checkmark \\
Test functions names & $\times$ & \checkmark \\
\bottomrule
\end{tabular}
\end{table}

%% file: appx/performance_by_donor_type.tex
\section{Models Performance by Donor Repository Type}
\label{section:performance-by-donor}

We analyzed the performance of models across different repository sources to investigate potential biases in our benchmark. The \benchmark{} dataset comprises 100 tasks derived from 34 distinct base repositories, with 66 examples originating from open-source repositories and 34 from custom-developed sources. While each task underwent substantial engineering modifications regardless of its origin, we sought to determine whether tasks derived from open-source repositories might be less difficult compared to custom-developed ones due to models being potentially trained on the open-source data.

As shown in Table~\ref{tab:model-metrics-by-source} for the majority of models the tasks derived from open-source repositories are equally or even more challenging. One important consideration is that randomness in the generation process and limited group sizes (34 tasks in the smallest group) suggest caution in the results interpretation.

\begin{table}[H]
\centering
\caption{Correctness of proprietary and open-source LLMs on the whole benchmark and separately for tasks created from open-source and from custom created repositories. Models are grouped by source type and provider, and sorted from most recent to oldest variant within each group. \textsuperscript{\dag} indicates LLMs executed with reasoning capabilities. \textsuperscript{\ding{61}} Claude 3.7 was executed without extended reasoning. Top results are in \textbf{bold}.}
\label{tab:model-metrics-by-source}
\resizebox{\textwidth}{!}{
\begin{tabular}{lllccc}
\toprule
\textbf{Source} & \textbf{Company} & \textbf{Model} & \multicolumn{3}{c}{\textbf{Correctness}} \\
\cmidrule(lr){4-6}
 & & & \textbf{All Tasks} & \textbf{Custom Derived Tasks} & \textbf{Open-Source Derived Tasks} \\
\midrule
\multirow{12}{*}{\rotatebox[origin=c]{90}{\textbf{Proprietary}}}
& Anthropic & Claude 3.7 Sonnet \textsuperscript{\ding{61}} & \textbf{34\%} & 35\% & \textbf{33\%} \\
&   & Claude 3.5 Sonnet V2 & 32\% & 32\% & 32\% \\
&   & Claude 3.5 Sonnet & 29\% & 32\% & 27\% \\
&   & Claude 3 Haiku & 8\% & 12\% & 6\% \\
\cmidrule(lr){2-6}
& Google & Gemini 2.5 Pro (03-25 Preview) \textsuperscript{\dag} & 29\% & \textbf{44\%} & 21\% \\
&   & Gemini 2.5 Pro (05-06 I/O Edition Preview) \textsuperscript{\dag} & 29\% & 38\% & 24\% \\
&   & Gemini 2.0 Flash Lite & 5\% & 6\% & 5\% \\
&   & Gemini 2.0 Flash & 4\% & 0\% & 6\% \\
\cmidrule(lr){2-6}
& OpenAI & OpenAI o3 \textsuperscript{\dag} & 23\% & 29\% & 20\% \\
&   & OpenAI o4-mini \textsuperscript{\dag} & 23\% & 26\% & 21\% \\
&   & GPT-4.1 & 18\% & 18\% & 18\% \\
&   & GPT-4o Mini & 4\% & 9\% & 2\% \\
\midrule
\multirow{8}{*}{\rotatebox[origin=c]{90}{\textbf{Open-Source}}}
& DeepSeek & DeepSeek R1 \textsuperscript{\dag} & 24\% & 26\% & 23\% \\
\cmidrule(lr){2-6}
& Mistral & Mistral Large & 14\% & 12\% & 15\% \\
&   & Codestral & 9\% & 15\% & 6\% \\
\cmidrule(lr){2-6}
& Meta & LLaMA 3.1 405B Instruct & 9\% & 12\% & 8\% \\
&   & LLaMA 3.1 70B Instruct & 3\% & 6\% & 2\% \\
&   & LLaMA 4 Maverick 17B Instruct & 8\% & 9\% & 8\% \\
&   & LLaMA 4 Scout 17B Instruct & 2\% & 3\% & 2\% \\
\cmidrule(lr){2-6}
& Alibaba & Qwen2.5 72B Instruct Turbo & 0\% & 0\% & 0\% \\
\bottomrule
\end{tabular}
}
\end{table}

%% file: appx/consistency_analysis.tex
\section{Consistency Analysis}
\label{section:consistency-analysis}

We quantified consistency using a "Success Consistency" metric. For each model and example, the "Task Success Consistency Gap" was calculated as the difference between the best and worst attempt correctness:

\begin{equation}
\text{Task Success Consistency Gap} = \max_{i \in \text{attempts}}(\text{correctness}_i) - \min_{i \in \text{attempts}}(\text{correctness}_i)
\end{equation}

Success Consistency is then defined as:

\begin{equation}
\text{Success Consistency} = 1 - \text{Task Success Consistency Gap}
\end{equation}

This value is averaged over all examples for each model. A value of 1 means perfectly consistent performance across attempts (either always failing or always succeeding). A value of 0 indicates maximum inconsistency (some attempts succeed while others fail).

As indicated in Section~\ref{sec:results} all LLMs demonstrate substantial improvement when given multiple opportunities to solve a task. Gemini 2.5 Pro achieves the highest pass@5 rate, indicating that for 47\% of tasks, at least one of five attempts produces a fully correct solution. This represents a considerable improvement over its single-attempt performance. DeepSeek R1 and OpenAI o3 show similar patterns, with pass@5 rates consistently higher than pass@1. Claude 3.7 Sonnet, however, shows higher average correctness of the results. It explains how Claude 3.7 Sonnet with its slightly lower pass@5 of 41\% achieves the top position in one-trial benchmarking. 

\input{figures/correctness-vs-consistency}

Figure~\ref{fig:corr-vs-consist} illustrates the relationship between average correctness across trials and success consistency for tested models. We observed that higher-performing on average Claude 3.7 Sonnet demonstrated better consistency, while Gemini 2.5 Pro, DeepSeek R1 and OpenAI o3 are not as consistent in providing correct results on this dataset. 

%% file: figures/correctness-vs-consistency.tex
\begin{figure}[H]
\centering
\begin{minipage}{\textwidth}
    \centering
    \includegraphics[width=0.9\textwidth]{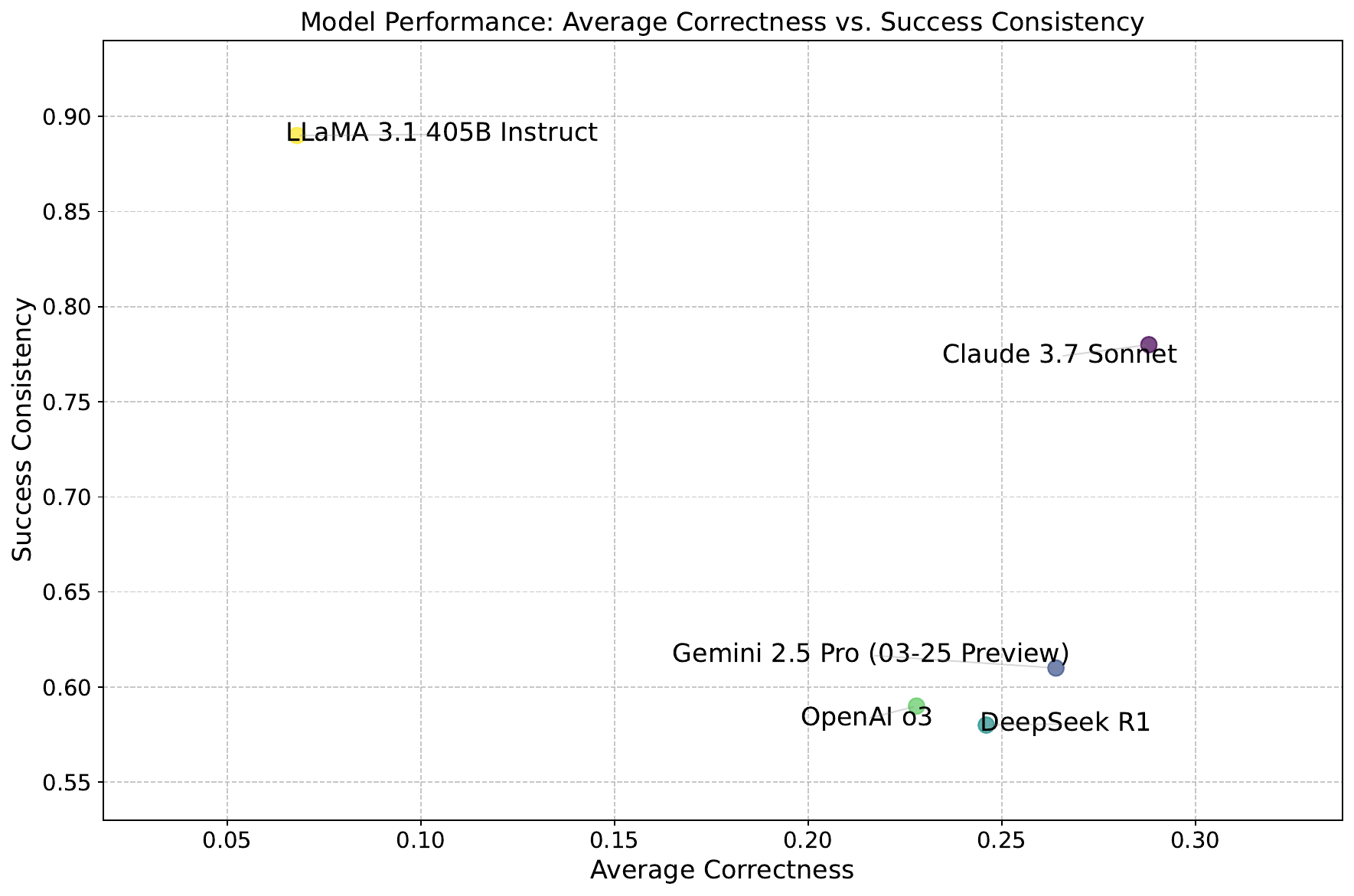}
\end{minipage}
\vspace{0.5cm} 
\vspace*{-2ex}
\caption{Average Correctness vs Consistency. Models positioned higher on average correctness, notably Claude 3.7 Sonnet, tend to demonstrate greater consistency, providing similar outcomes across repeated attempts. In contrast, models such as Gemini 2.5 Pro, DeepSeek R1, and OpenAI o3, despite achieving significant improvements when allowed multiple attempts, exhibit less consistency, indicating more variability in their success across trials.}
\label{fig:corr-vs-consist}
\end{figure}

%% file: appx/error-distribution.tex
\section{Error Type Distribution}
\label{section:error-dist}

Figure~\ref{fig:error-transition} illustrates how error distributions change when models are given a second attempt with feedback (high verbosity configuration) from their first attempt. All models benefit from the second attempt, improving both on syntax and logical errors. However, a subset of LLMs without reasoning, namely GPT-4.1, Mistral Large and Llama 4 Maverick 17B Instruct, retain the same percentage of logical errors, mostly correcting the syntax ones. Such a behavior indicates that even high verbosity level is insufficient to address these more complex, implementation-specific failures. On the other hand, Claude Sonnet models, as well as Gemini 2.5 and DeepSeek R1, that further leverage their reasoning capabilities, showcase the considerable improvement on this matter as a prove of better analysis of underlying infrastructure dependencies and relationships.

\clearpage

\input{figures/error_transition}

%% file: figures/error_transition.tex
\begin{figure}[H]
\centering
\includegraphics[width=0.85\textwidth]{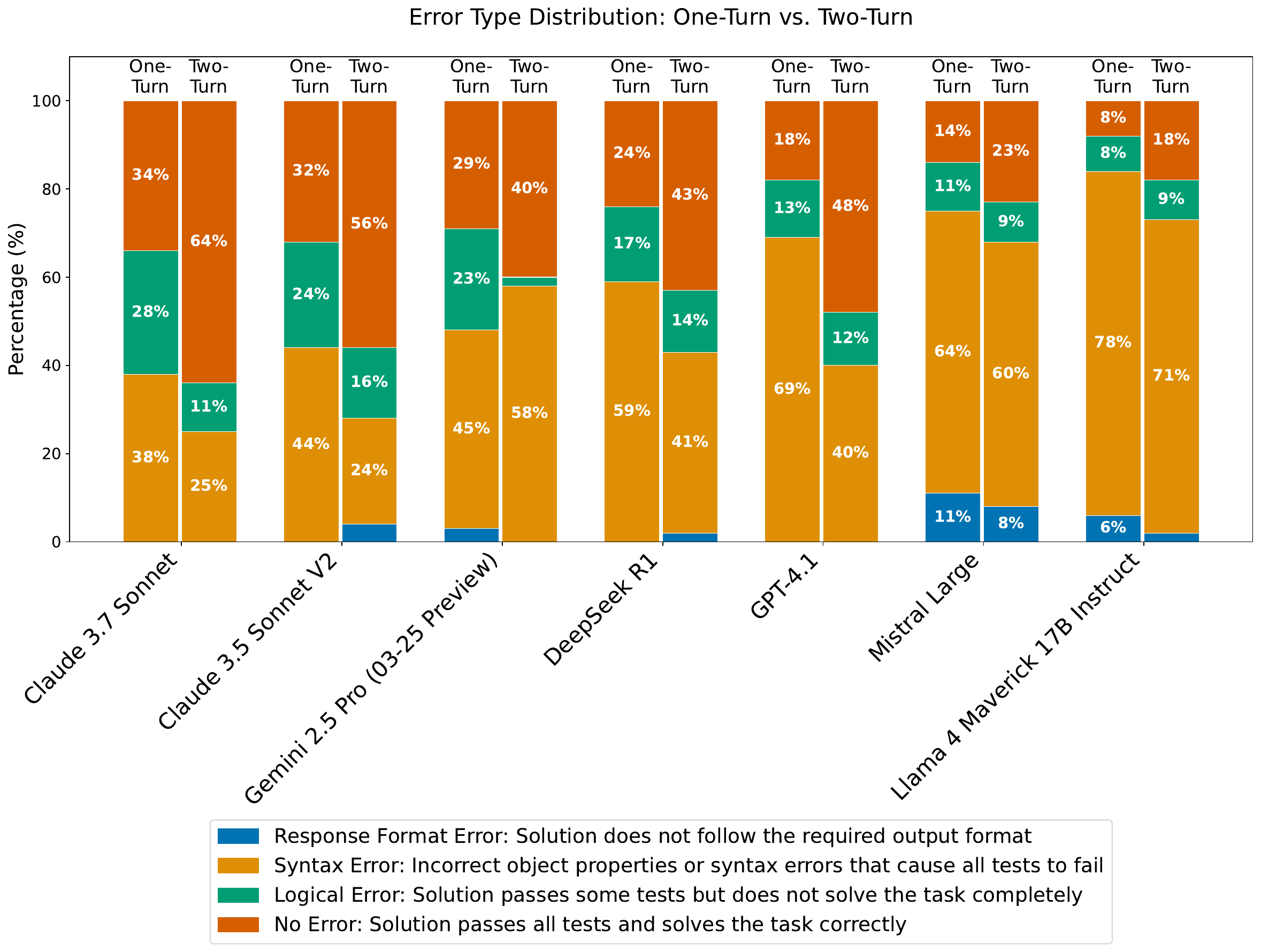}

\caption{Error type distribution comparison between one-turn and two-turn (without RAG, high verbosity) approaches. All models show improved performance in the two-turn approach. The Claude family models demonstrate balanced improvement by reducing both syntax and logical errors. Gemini model primarily addresses logical errors in its second attempt, while GPT-4.1 and DeepSeek R1 shows substantial reduction in syntax errors. While most examples benefit from the two-turn approach, a small percentage show regression due to the inherent randomness in the generation process.}
\label{fig:error-transition}
\end{figure}